\documentclass[prd,preprint,nofootinbib]{revtex4-1} 
\usepackage{latexsym,verbatim}
\usepackage{amssymb}
\usepackage{amsfonts}
\usepackage{amsmath,color}
\usepackage[draft]{graphicx}
\usepackage{bm}
\usepackage[utf8]{inputenc}
\usepackage[T1]{fontenc}

\usepackage{hyperref}

\hypersetup{
    colorlinks=true,
    linkcolor=red,
    citecolor=red,
    filecolor=magenta,      
    urlcolor=cyan,
    pdftitle={Stationary rotating and axially symmetric dust systems as peculiar General Relativistic objects},
    pdfpagemode=FullScreen,}

\def\mb#1{\mathbf{#1}}

\def\ber{\begin{eqnarray}}
\def\eer{\end{eqnarray}}
\def\beq{\begin{equation}}
\def\eeq{\end{equation}}

\def\ed{\end{document}}

\makeatletter
\let\jnl@style=\rm
\def\ref@jnl#1{{\jnl@style#1}}

\def\aj{\ref@jnl{AJ}}                   % Astronomical Journal
\def\actaa{\ref@jnl{Acta Astron.}}      % Acta Astronomica
\def\araa{\ref@jnl{ARA\&A}}             % Annual Review of Astron and Astrophys
\def\apj{\ref@jnl{ApJ}}                 % Astrophysical Journal
\def\apjl{\ref@jnl{ApJ}}                % Astrophysical Journal, Letters
\def\apjs{\ref@jnl{ApJS}}               % Astrophysical Journal, Supplement
\def\ao{\ref@jnl{Appl.~Opt.}}           % Applied Optics
\def\apss{\ref@jnl{Ap\&SS}}             % Astrophysics and Space Science
\def\aap{\ref@jnl{A\&A}}                % Astronomy and Astrophysics
\def\aapr{\ref@jnl{A\&A~Rev.}}          % Astronomy and Astrophysics Reviews
\def\aaps{\ref@jnl{A\&AS}}              % Astronomy and Astrophysics, Supplement
\def\azh{\ref@jnl{AZh}}                 % Astronomicheskii Zhurnal
\def\baas{\ref@jnl{BAAS}}               % Bulletin of the AAS
\def\bac{\ref@jnl{Bull. astr. Inst. Czechosl.}}
\def\caa{\ref@jnl{Chinese Astron. Astrophys.}}
\def\cjaa{\ref@jnl{Chinese J. Astron. Astrophys.}}
\def\icarus{\ref@jnl{Icarus}}           % Icarus
\def\jcap{\ref@jnl{J. Cosmology Astropart. Phys.}}
\def\jrasc{\ref@jnl{JRASC}}             % Journal of the RAS of Canada
\def\memras{\ref@jnl{MmRAS}}            % Memoirs of the RAS
\def\mnras{\ref@jnl{MNRAS}}             % Monthly Notices of the RAS
\def\na{\ref@jnl{New A}}                % New Astronomy
\def\nar{\ref@jnl{New A Rev.}}          % New Astronomy Review
\def\pra{\ref@jnl{Phys.~Rev.~A}}        % Physical Review A: General Physics
\def\prb{\ref@jnl{Phys.~Rev.~B}}        % Physical Review B: Solid State
\def\prc{\ref@jnl{Phys.~Rev.~C}}        % Physical Review C
\def\prd{\ref@jnl{Phys.~Rev.~D}}        % Physical Review D
\def\pre{\ref@jnl{Phys.~Rev.~E}}        % Physical Review E
\def\prl{\ref@jnl{Phys.~Rev.~Lett.}}    % Physical Review Letters
\def\pasa{\ref@jnl{PASA}}               % Publications of the Astron. Soc. of Australia
\def\pasp{\ref@jnl{PASP}}               % Publications of the ASP
\def\pasj{\ref@jnl{PASJ}}               % Publications of the ASJ
\def\rmxaa{\ref@jnl{Rev. Mexicana Astron. Astrofis.}}%
\def\qjras{\ref@jnl{QJRAS}}             % Quarterly Journal of the RAS
\def\skytel{\ref@jnl{S\&T}}             % Sky and Telescope
\def\solphys{\ref@jnl{Sol.~Phys.}}      % Solar Physics
\def\sovast{\ref@jnl{Soviet~Ast.}}      % Soviet Astronomy
\def\ssr{\ref@jnl{Space~Sci.~Rev.}}     % Space Science Reviews
\def\zap{\ref@jnl{ZAp}}                 % Zeitschrift fuer Astrophysik
\def\nat{\ref@jnl{Nature}}              % Nature
\def\iaucirc{\ref@jnl{IAU~Circ.}}       % IAU Cirulars
\def\aplett{\ref@jnl{Astrophys.~Lett.}} % Astrophysics Letters
\def\apspr{\ref@jnl{Astrophys.~Space~Phys.~Res.}}
\def\bain{\ref@jnl{Bull.~Astron.~Inst.~Netherlands}}
\def\fcp{\ref@jnl{Fund.~Cosmic~Phys.}}  % Fundamental Cosmic Physics
\def\gca{\ref@jnl{Geochim.~Cosmochim.~Acta}}   % Geochimica Cosmochimica Acta
\def\grl{\ref@jnl{Geophys.~Res.~Lett.}} % Geophysics Research Letters
\def\jcp{\ref@jnl{J.~Chem.~Phys.}}      % Journal of Chemical Physics
\def\jgr{\ref@jnl{J.~Geophys.~Res.}}    % Journal of Geophysics Research
\def\jqsrt{\ref@jnl{J.~Quant.~Spec.~Radiat.~Transf.}}
\def\memsai{\ref@jnl{Mem.~Soc.~Astron.~Italiana}}
\def\nphysa{\ref@jnl{Nucl.~Phys.~A}}   % Nuclear Physics A
\def\physrep{\ref@jnl{Phys.~Rep.}}   % Physics Reports
\def\physscr{\ref@jnl{Phys.~Scr}}   % Physica Scripta
\def\planss{\ref@jnl{Planet.~Space~Sci.}}   % Planetary Space Science
\def\procspie{\ref@jnl{Proc.~SPIE}}   % Proceedings of the SPIE

\makeatother

\begin{document}

\author{Matteo Luca Ruggiero}
\email{matteoluca.ruggiero@unito.it}
\affiliation{Dipartimento di Matematica ``G.Peano'', Universit\`a degli studi di Torino, Via Carlo Alberto 10, 10123 Torino, Italy}
\affiliation{INFN - LNL , Viale dell'Universit\`a 2, 35020 Legnaro (PD), Italy}

\date{\today}

\title{Stationary rotating and axially symmetric dust systems as peculiar General Relativistic objects}

\begin{abstract}
We study an exact solution of Einstein's equations describing a self-gravitating system, made of dust, distributed with axial symmetry and in stationary rotation, and we prove that this type of system has no Newtonian analogue. In a low-energy limit, its existence depends on the solution of a Grad-Shafranov equation in vacuum which can be interpreted as a Laplace equation for the toroidal component of the gravitomagnetic potential; in particular, in this system the relativistic rotational effects are of the order of magnitude of Newtonian ones. We therefore argue that this exact solution should contain singularities and discuss the possible consequences of using such a system as simplified model for galactic dynamics.
\end{abstract}

\maketitle

%%------------------------Section-------------------------
\section{Introduction}\label{sec:intro}
%%------------------------Section-------------------------

General Relativity (GR) is the best model that we have to describe gravitational interactions: one century after its birth, we know that it passed with great success numerous tests and helped to greatly improve our knowledge of the near and far Universe \cite{Will:2014kxa}. The Einsteinian picture drastically changed our understanding of the structure of spacetime, however GR effects can be often considered small corrections with respect to the Newtonian theory of gravitation, at least in regions where the gravitational field is weak and the speeds are small if compared to the speed of light, which is reasonably true in the terrestrial environment and in the Solar System. Nonetheless, extreme astrophysical events exist in which spacetime is greatly deformed by the presence of very compact objects that are fast moving or rotating. In these cases, new phenomena arise which do not possess at all a Newtonian analogue: just to mention few of them,  we can think of the existence of neutron stars, black holes and of the emission of gravitational waves.

Actually, also in condition of weak-gravitational field there are GR effects without Newtonian analogue: this is the case of the so-called gravitomagnetic effects \cite{Ruggiero:2023ker} which, roughly speaking, are determined by mass currents. As a matter of fact, it is not  true that GR effects are always (much) smaller than the corresponding Newtonian ones, since the latter at times simply do not exist; but even if they do exist, the situation is not always straightforward. In fact, if we consider the bending of a light a ray by a source like the Sun, we know that it can be calculated using a Newtonian approach, but the result differs by a factor 2 from the general relativistic one \cite{rindler2006relativity}: hence, both the Newtonian and the GR effect are of the same order of magnitude.

The purpose of this paper is to discuss another situation where, surprisingly enough, GR and Newtonian effects are expected to be of the same order of magnitude and, in addition, the very existence of the system under consideration could not be possible in a classical, i.e. Newtonian, framework. The motivations derive from this simple question: ``Do there exist general relativistic self-gravitating systems, made of dust, in stationary and axially symmetric rotation?''. After analyzing the question in the context of the exact solutions of Einstein's equations, we suggest that if such systems can be used as a model for a galaxy, its dynamics, i.e. the rotation curves, are also influenced by peculiar relativistic effects.

%%------------------------Section-------------------------
\section{The Exact Solution}\label{sec:sol}
%%------------------------Section-------------------------

We use cylindrical coordinates  $\left(ct,\phi,r,z \right)$ and the signature is $(-1,1,1,1)$; due to the symmetry of the system, we know that matter is allowed to flow along the Killing vectors $\partial_t$ and $\partial_\phi$: as a consequence,  all functions considered will depend on the coordinates $(r,z)$ only. Accordingly, we may write the energy momentum tensor $T^{\mu\nu} = \rho u^{\mu} u^{\nu}$, where $\rho$ is the energy density ($\rho=\rho_{m}c^{2}$, where $\rho_{m}$ is the matter density) and $u^{\mu}$ is the fluid four-velocity.  Given these symmetries and matter distribution, Einstein's equations can be integrated up to quadratures, using techniques that can be traced to the work of \citet{Geroch:1970nt},\citet{Geroch:1972yt}, \citet{HansenWinicour1}, \citet{HansenWinicour2}, \citet{Winicour}: a summary of the approaches to these kinds of exact solutions can be found in the textbook by \citet{kramerstephani}, where it is shown that the solution of Einstein's equations is completely determined by the choice of a negative function $H(\eta)$, on which the physical properties depend: the meaning of $\eta$ will be clarified below. Accordingly, the fluid velocity can be written as $u^{\mu}= \frac{1}{\sqrt{-H}} \left(1,\Omega,0,0 \right)$, where $\Omega= \frac{d\phi}{dt}= \frac{u^{\phi}}{u^{t}}$ is the angular velocity of the fluid as seen by observers at rest with respect to the given set of coordinates. The function $H(\eta)$ depends on the existence of the  auxiliary function $\mathcal{F}(\eta)$ \footnote{We use the following notation: for any function of one argument, like $H(\eta)$, with a prime we mean the derivative with respect to its argument; in addition, we use a comma to indicate partial derivative with respect to  a given coordinate.}
\begin{equation}
\label{def v}
	\mathcal{F}=2\eta+r^2\int\frac{H'}{H}\frac{d\eta}{\eta}-\int\frac{H'}{H}\eta d\eta,
\end{equation}%
which needs to identically satisfy the equation
\beq
\mathcal{F}_{,rr}-\frac{1}{r}\mathcal{F}_{,r}+\mathcal{F}_{,zz}=0. \label{harm}
\eeq
Once that $H(\eta),\mathcal F(\eta)$ have been determined, it is possible to obtain the fluid angular velocity
\beq
\Omega=\frac{1}{2}\int H'\frac{d\eta}{\eta} \label{eq:omega}. 
\eeq
In summary, the metric components read
\begin{eqnarray}
	g_{tt}&=&\frac{(H-\eta\Omega)^2-r^2\Omega^2}{H}, \label{components1} \\
	g_{t\phi}&=&\frac{\eta^2-r^2}{(-H)}\Omega+\eta, \label{components2} \\
	g_{\phi\phi}&=&\frac{r^2-\eta^2}{(-H)} \label{components3}.
\end{eqnarray}
and the remaining metric components $g_{zz}=g_{rr}=:e^{\Psi}$ can be obtained using the following equations 
\begin{align}
    \Psi_{,r} =& \frac{1}{2r} \left[ (g_{tt})_{,r} (g_{\phi\phi})_{,r}-(g_{tt})_{,z} (g_{\phi\phi})_{,z} - ((g_{t\phi})_{,r} )^2+((g_{t\phi})_{,z} )^2 \right]     \label{mur},\\
    \Psi_{,z}= & \frac{1}{2r} \left[ (g_{tt})_{,z} (g_{\phi\phi})_{,r} + (g_{tt})_{,r} (g_{\phi\phi})_{,z} -  2 (g_{t\phi})_{,r} (g_{t\phi})_{,z}  \right] \label{muz}.
\end{align}
Moreover, the energy density is given by
\begin{equation}
\label{rho}
	8\pi G\rho=\frac{\eta^2 r^{-2}(2-\eta l)^2-r^2l^2}{4g_{rr}}\frac{\eta_{,r}^2+\eta_{,z}^2}{\eta^2},
\end{equation}
where  $\displaystyle l= \frac{H'}{H}$.\\

We can learn more about the meaning of this solution if we consider the Zero Angular Momentum Observers (ZAMO) \cite{zamo}: as we discussed in \cite{astesiano}, the metric can be written in the form
\begin{align}
    ds^2&= H\gamma^{2}{} c^{2}dt^2 - r^2 \frac{1}{H\gamma^{2}}\left(d\phi-\chi dt \right)^2+ e^{\Psi}\left(dr^2+dz^2\right), \label{IR} 
\end{align}
where $ \gamma=\frac{1}{\sqrt{1-\frac{v^{2}}{c^{2}}}}$, being $v$ the velocity of the dust fluid as measured by the ZAMO, and $ \chi \equiv - \frac{g_{t\phi}}{g_{\phi\phi}}=  \frac{H \eta}{(r^2-\eta^2)}+\Omega$ is the angular velocity of the ZAMO as seen by   asymptotic inertial observers   at infinity. In addition, $\eta=vr$, so this function is related to the angular momentum per unit mass of a dust element. It is possible to get the following relation
\begin{align}
    r \Omega = r \chi  -v \gamma^2\, H \label{Chi}
\end{align}
between the coordinate velocity $r\Omega$ of the dust, its corresponding ZAMO expression $v$ and the ZAMO's velocity $r\chi$. 

The metric (\ref{IR}) is  \textit{non time-orthogonal}, because $g_{0i} \neq 0$: this is an expected feature, since these off-diagonal terms are generally related to the rotational features of the reference frame and to the rotation of the sources of the gravitational field \cite{Ruggiero:2023ker}. In particular, from $g_{0i}$ it possible to formally introduce a gravitomagnetic potential
\beq
A_i=-c^2\frac{g_{0i}}{2} \label{eq:defAphi}
\eeq  
which, in the weak-field and slow-motion approximation, enables to describe the motion of free test particles in terms of the action of a Lorentz-like force equation,  exploiting the gravitoelectromagnetic analogy \cite{Ruggiero:2002hz,mashhoon03,Ruggiero:2023ker}. In our case the gravitomagnetic effects are related to the function $\chi$, since $g_{0\phi}=\frac{r^{2}\chi}{H\gamma^{2}}$ and $\mb A= A_{\phi}\mb e_{\phi}$.

%%------------------------Section-------------------------
\section{The Equilibrium Conditions}\label{sec:equil}
%%------------------------Section-------------------------

The exact solution considered describes the motion of a dust fluid; from  the conservation law   of the energy-momentum tensor $T^{\mu\nu}_{\ \ ;\nu}=0$, we deduce that  the dust elements are in geodesic motions, which \textit{by construction}   are circular trajectories in planes at constant $z$. Let us see a first consequence of this hypothesis. For simplicity, we define the function $a=-H\gamma^{2}$ in the metric (\ref{IR}), and then we write the Lagrangian
\beq
\mathcal L=\frac 1 2 \left[\left(-a+\frac{r^{2}}{a}\chi^{2} \right)c^{2}-\frac{2r^{2}\chi}{a}\dot \phi+\frac{r^{2}}{a}\dot \phi^{2}+ e^{\Psi}\left(\dot r^{2}+\dot z^{2}\right) \right],\label{eq:L1}
\eeq
where dot means derivative with respect  to the coordinate time. Now, we are interested in the components of the geodesics in the $z$ direction: we get $\frac{\partial \mathcal L}{\partial \dot z}=e^{\Psi}\dot z$ and, on setting $z=\mathrm{const}$, from the Euler-Lagrange equation we get $\frac{\partial \mathcal L}{\partial  z}=0$, or
\beq
\frac{\partial}{\partial z} \left(-a+\frac{r^{2}}{a}\chi^{2} \right)c^{2}+\frac{\partial}{\partial z} \left(-\frac{2\chi}{a} \right) r^{2} \dot \phi+\frac{\partial}{\partial z} \left(\frac{1}{a} \right)r^{2}\dot\phi^{2}=0. \label{eq:L2}
\eeq 
If we suppose that $\chi=0$, we get
\beq
\frac{\partial a}{\partial z} \left(c^{2}+\frac{r^{2}\dot \phi^{2}}{a^{2}} \right)=0.  \label{eq:condz1}
\eeq
So, in this case circular geodesics at $z=\mathrm{const}$ are realizable only if the system has \textit{cylindrical symmetry}, which  means that it is not possible to obtain a compact structure.  Actually, this is what happens in Newtonian gravity, where no compact or finite dust object can exist, as  \citet{bonnor1977rotating} pointed out. 

Until now, we made no assumptions on the nature of the system we are considering. If we suppose that we refer to an actual physical system, it is reasonable to  expect that this solution can be used to describe some low-energy limits and, in this condition, the exact metric (\ref{IR}) can be expanded in negative powers of $c$, as it is customary in the post-Newtonian development  \cite{astesiano} . Accordingly, we may write $a=1-\frac{2U}{c^{2}}+O(c^{-4})$, where $U$ is the gravitoelectric or Newtonian potential\footnote{Notice that $U$ is defined in analogy with electromagnetism and differs by a minus sign from the standard definition of  the Newtonian potential.}; to simplify the results, we introduce the function $\psi=\chi r^{2}$. Now, we consider the Euler-Lagrange equations for the coordinates $r,z$ and, by hypothesis, we set $z=\mathrm{const}, r=\mathrm{const}$ to describe the geodesic motions of the dust fluid.  We get
\begin{eqnarray}
0 & = & \frac{\partial U}{\partial r}+\frac{\psi}{r}\frac{\partial}{\partial r}\left(\frac \psi r \right)-\frac{\partial \psi}{\partial r} \dot \phi +r\dot \phi^{2} \label{eq:geor} \\
0 & = & \frac{\partial U}{\partial z}+\frac{\psi}{r^{2}}\frac{\partial \psi}{\partial z}-\frac{\partial \psi}{\partial z} \dot \phi \label{eq:geoz}
\end{eqnarray}
We see from the above equations that, in order to get the equilibrium for the geodesic motions, the effects determined by $\psi$ needs to be of the same order as the Newtonian ones.

In addition, using relations (\ref{eq:omega}) and (\ref{Chi}), we can calculate the expression of the function $\mathcal F$ from (\ref{def v}), and we get $\mathcal F=-2\psi$. Accordingly, the function $\psi$ satisfies the equation
\beq
{\psi}_{,rr}-\frac{1}{r}\psi_{,r}+{\psi}_{,zz}=0. \label{eq:Aharm}
\eeq
which is in the form of the homogenous Grad-Shafranov equation \cite{grad1958hydromagnetic,shafranov1958magnetohydrodynamical}. The latter equation is often used to describe the equilibrium of a two dimensional plasma, in magnetohydrodynamics; in particular, it is easy to show that  using the definition of the gravitomagnetic potential (\ref{eq:defAphi}), the above equation (\ref{eq:Aharm}) can be regarded as a Laplace equation for the gravitomagnetic vector potential $\mb A=A_{\phi}\mb e_{\phi}$: 
\beq
\nabla^{2} \mb A=0 \label{eq:laplace}
\eeq
If we compare the above equation with the corresponding one  obtained when we consider the weak-field and slow-motion approximation of Einstein's 
equations \cite{Astesiano:2022ghr}
\beq
 \nabla^{2} \mb A = -\frac{8\pi G}{c}\, \mb j, \label{eq:laplacej}
\eeq
where $\mb j$ is the mass-energy currents,  we see that the gravitomagnetic potential determined by $\psi$ \textit{is not originated by the local mass distribution}, rather its sources should be located elsewhere and, as we are going to show below, they have a singular behaviour at infinity or along the siymmetry axis.  

To avoid misunderstandings, it is important to give the right meaning to words. Even if there are various gravitoelectromagnetic analogies that arise in GR \cite{Ruggiero:2023ker},  gravitomagnetic effects are generally understood as the solutions of Eq. (\ref{eq:laplacej}), while the gravitoelectric ones are the solutions
of the corresponding equation for the gravitoelectric or Newtoninan potential. Properly,  in the  Newtonian limit, $U$ reduces to the Newtonian gravitational potential, while $A_{i} = O(c^{-1}$) \cite{mashhoon03}. 

But what we are focusing on here is different: in fact  the solutions of Eqs (\ref{eq:Aharm}) or (\ref{eq:laplace})  by no means  go to zero as far as $c\rightarrow \infty$. From now on, we will call them \textit{rotation effects}  (or \textit{homogenous solutions} as done   previous works \cite{Astesiano:2021ren,astesiano,Astesiano:2022ghr}) to distinguish them from the popular gravitomagnetic ones.\\

The fact that the Grad–Shafranov equation in vacuum coincides with the Laplace equation for the toroidal component of the vector potential  \cite{crisanti,cesarano} suggests that the solutions can be found in analogy with electromagnetism. For instance, since $\displaystyle \mb A=\frac{\mb m \wedge \mb x}{|\mb x|^{3}}$ is  the solution of the Laplace equation describing the vector potential of a magnetic-dipole $\mb m$, we see that
\beq
\psi =\frac{m r^{2}}{\left(r^{2}+z^{2}\right)^{3/2}} \label{eq:psidipole}
\eeq  
{is a solution of the above equation (\ref{eq:Aharm}) and corresponds to the Bonnor's solution \cite{bonnor1977rotating}.} More generally speaking, it is possible to obtain the solution of the above equation (\ref{eq:Aharm}) as a multipole expansion (see e.g. \citet{multipole1,multipole2}): using spherical coordinates $R,\theta,\varphi$, the solutions are in the form
\beq
\psi(R,\theta)=\sum_{n=2}^{\infty}\left(\alpha_{n}R^{n}+\beta_{n}R^{1-n} \right) \sin \theta P^{1}_{n-1}(\cos \theta) \label{eq:solsph2} 
\eeq
where $P^{1}_{n-1}(\cos \theta)$ are the Legendre functions, and $\alpha_{n},\beta_{n}$ are arbitrary constants. Notice that the solutions which multiply $\alpha_{n}$ are regular along the symmetry axis, while the others are regular at  infinity. In particular, the solution (\ref{eq:psidipole}) corresponds to  $\alpha_{2}=0$ and $\beta_{2}=m$ and the other terms are null. 

We remark that if suppose that our system has a finite extension,  the solutions that are regular at the origin do not necessarily give singularities at infinity, because it is expected that the internal solution described by (\ref{IR}) should be matched to an external solution that extends to infinity.

%%------------------------Section-------------------------
\section{Discussion and Conclusions}\label{sec:discconc}
%%------------------------Section-------------------------

We considered a self-gravitating system, made of an axially symmetric dust fluid in stationary rotation: the metric elements of the corresponding exact solution of Einstein's equations are given by Eqs. (\ref{components1})-(\ref{muz}). These elements are completely determined by the choice of the negative function $H(\eta)$, taking into account  the  auxiliary function $\mathcal{F}(\eta)$ which satisfies the  condition expressed by Eq. (\ref{harm}). Since, by hypothesis, the system is made of dust particles, their motion is geodesic: accordingly, the solution of the geodesic equations must give circular spatial trajectories at constant $z$ coordinate. 

A first point that needs to be stressed is that the very existence of this system rests on the presence of the rotation effects determined by the solution of Eq. (\ref{eq:Aharm}) in the low-energy limit, or of Eq. (\ref{harm}) in the exact solution: in fact, if they were absent, the system would be cylindrical symmetric, i.e. with infinite extension along the symmetry axis. This is what happens in Newtonian gravity, where it is impossible to build a  limited system, stationary rotating  with axial symmetry: so, the system that we are considering is peculiar since it has no  Newtonian analogue. 

A second important point is that an inspection of the geodesic equation (\ref{eq:geoz}) reveals that  to have an equilibrium along the symmetry axis, the rotation effects determined by  $\psi$  and deriving from the off-diagonals terms in the spacetime metric, cannot be negligible with respect to the Newtonian ones, represented by $U$.

The rotation effects  stem from the solution of the  vacumm Grad-Shafranov  equation which can be interpreted as a Laplace equation for the toroidal component of the gravitomagnetic potential. Consequently, what we have shown suggests that, if they exist, these exact solutions of Einstein's equations should have singularities. This is not surprising: in fact, a particular case of this class is represented by the Balasin  and Grumiller  solution \cite{Balasin:2006cg}, which describes a rigidly rotating (i.e. $\Omega=\mathrm{const}$) dust \cite{Astesiano:2022gph}. A recent analysis by \citet{costaBG} shows that this solution contains singularities along the axis, namely a pair of NUT   rods and a cosmic string; we remember that   a Newman–Unti–Tamburino (NUT) spacetime is a solution of Einstein's equations that generlises the Schwarzschild solution since, in addition to the mass parameter, it contains a second parameter, the so-called NUT charge, that can be interpreted as gravitomagnetic monopole (see e.g. \citet{perlick} and references therein).

Seemingly, the solution of Einstein's equation describing a self-gravitating system, made of an axially symmetric dust fluid in stationary rotation, requires a vacuum solution for the rotation term $\psi$. Notice that, as discussed  by  \citet{Astesiano:2022ghr}, in a low-energy limit, these vacuum solutions become sources of the Poisson equation for the Newtonian potential: 
\beq
 \left[\nabla^2 U+ \frac{(\partial_{z}\psi)^2+(\partial_r\psi-2 \frac{\psi}{r})^2}{2 r^2}\right]=-4\pi G\rho_{m}\\  \label{eq:poissonmod}
\eeq
Actually, this is not surprising since the same happens for  the exact axially symmetric solutions of Einstein equations in vacuum (see e.g. \citet{treves,bonnor_exact}). Differently speaking, our approach  naturally suggests that spacetime curvature, through the rotation term $\psi$, modifies the interplay between the sources of the gravitational field and the Newtonian potential $U$:  the key point is that these additional sources are not necessary small. The $\psi$ term contributes with an effective matter density in the form
\beq
\rho_{\psi}=\frac{1}{4\pi G}\left(\frac{(\partial_{z}\psi)^2+(\partial_r\psi-2 \frac{\psi}{r})^2}{2 r^2} \right).  \label{eq:rhopsi1}
\eeq
In particular, we get for the dipole solution  (\ref{eq:psidipole}), $\rho_{\psi}=\frac{1}{8\pi G}\left( \frac{9mr^{2}}{(r^{2}+z^{2})^{4}} \right)$, which is rapidily increasing at the origin. On the other hand, if we consider a solution in the form $\psi=\alpha_{3}zr^{2}$, we get $\rho_{\psi}=\frac{1}{8\pi G}\left(\alpha_{3} r^{2} \right)$, which is smooth at the origin and has cylindrical symmetry. 

Eq. (\ref{eq:poissonmod}) can be  interpreted in a Machian sense, since the  state of the system, i.e. its rotation with respect to asymptotical inertial observers, determines the local effective mass distribution  which is the source of the Poisson equation.

Our analysis is quite general and does not depend on the choice of a specific system, which can be defined only when a given mass distribution is taken into account. Conversely, it shows that the existence of such a system is determined by rotation effects which are of the same order of magnitude of Newtonian ones: in other words, this is a purely relativistic system, which cannot be studied in analogy with the Newtonian case, but only using the framework of General Relativity.

The simple question: ``are there any self-gravitating systems, made of dust, stationary rotating with axial symmetry?'' seemingly leads us to the key role of rotation effects, that are naturally incorporated in GR, but absent in pre-relativistic gravity.  We also emphasize that the system under consideration is by no means exotic, but it is made of the simplest kind of matter: dust.  We argue that the existence of this type of systems  can be intended as a new test of GR and we suggest that astrophysics is a  natural scenario to look for possible candidates.

{However, to ascertain the existence of such systems, it is crucial to note that, as of today, there is no known global solution to Einstein's equations that describes a rotating isolated matter distribution: the solution considered here is valid \textit{within} the source. This is primarily due to the absence of a systematic procedure for matching the internal solution  to the external one along an unknown surface. Consequently, global solutions remain elusive, with only limiting cases, such as rotating disks, being attainable \cite{Neugebauer93,Neugebauer:1996sz}. Nevertheless, it has been proposed that this system could be regarded as a model for a rotating cloud of dark matter, as dust gravitates without interacting with other particles: in particular, \citet{bambi} explored Bonnor's solution, as expressed in Eq. (\ref{eq:psidipole}), given its asymptotically flatness and the positivity of its energy density throughout. This renders it physically plausible, with the exception of a singularity at the center attributed to the diverging vorticity field of the dust fluid there \cite{Astesiano:2023emg}.\\
\indent Another possible application of such systems was considered in  previous works \cite{Astesiano:2021ren,astesiano,Astesiano:2022ghr}, where the relevance in the study of galactic dynamics was focused on. In doing so, it was assumed that a galaxy can  be modelled  as a rotating dust system:  the present analysis suggests that for such a model to exist it would have to contain singularities.}  

At this point, a clarification is important: when a galaxy is considered as a collisionless system, the distribution functions  (DF) that are solutions of the  Vlasov equation are considered (see e.g. \citet{binneytremaine}). This equation can be coupled to the Einstein equations by specifying the energy-momentum tensor in terms of the distribution function. However, it is possible to show (see \cite{rendall} and references therein) that the solutions of the Vlasov-Einstein system where the DF has a  distributional form are in one-to-one correspondence with dust solutions of the Einstein equations, where dust is meant to be a perfect fluid with zero pressure, which is what we consider.

It is significant to emphasize that the application of the solution considered in this and our previous papers to the galactic dynamic problem is different from the approach  proposed by \citet{Ludwig}, who considered the gravitomagnetic  effects originating from mass currents into the solution of Einstein equations in  weak-field and slow-motion approximation. In fact,  in this regime mass currents produce post-Newtonian effects and their impact is negligibly small with respect to the dominant Newtonian ones \cite{Ruggiero:2021lpf,Ciotti_2022}.

 A recent work by \citet{jan}  focuses on an exact  vacuum solution of GR equations, describing two rotating massive black holes of equal masses carrying opposite NUT charges along the symmetry axis;  in this work, the possibility is suggested that the flattening of the galactic rotation curves \cite{rubin1978extended,sofue2001rotation,sofuegalaxy} could be a consequence of these singular energy-momentum distributions, positioned along the rotation axis at a distance much larger than the visible spatial extent of the galaxy. In particular, the rotation effects are in the form of a dipole contribution like (\ref{eq:psidipole}). The  author  points out that his results are just  preliminary  but, in view of our analysis, they appear intriguing. 

We are aware that there is no guarantee that a galaxy can be described as a rotating dust fluid: however, if this can be done, at least for a very simplified model, our analysis suggests that singularities can play a role on its dynamics. It is relevant to point out that there are suggestions that collapsed objects could be described by a Kerr-Taub-NUT \cite{miller} spacetime, instead of a Kerr spacetime (see \citet{attuale1,attuale2} and references therein): in other words, the debate about the nature of the singularity hosted by a galaxy is indeed open. As a matter of fact, after one century of relativity, we learned that exact solutions of Einstein's equations must be taken seriously, even if they denote a strange behaviour: in fact, we have now a concrete evidence that a black hole exists at the center of a galaxy \cite{ETBH} and, accordingly,  the Schwarzschild or, more generally, the Kerr metric can be a faithful description of natural phenomena.

In conclusion, we have shown that the presence of  $\psi$ introduces a richer geometric structure, whose impact on the effective sources of the Newtonian potential $U$ is not trivial; furthermore the geodesic equations are greatly influenced by the presence of the rotation terms $\psi$. As a result, we expect rotational effects
can have a twofold impact on the system dynamics. In this regard, we emphasize once again that we do not want to claim that GR solutions can explain rotation curves without dark matter: 
rather, we suggest that there are hints that if a galaxy (or at least a limited region) can be modeled in this way, the curvature of spacetime may play a role on its dynamics. This fact should be studied to better understand the geometric structure and, if applicable, the impact of dark matter. {Or, 
alternatively, viewed from a different perspective,  this dust fluid could be a model for dark matter itself, as previously mentioned.}

In any case, we believe that these systems of self-gravitating dust in stationary rotation, due to their peculiar relativistic nature, deserve further attention to understand if they can be considered a model of real astrophysical objects.

\section*{Acknowledgments}

The author thanks Davide Astesiano for his collaboration on these topics, and Luca Ciotti for useful remarks; in addition, the author acknowledges the contribution of  the local research  project Modelli gravitazionali per lo studio dell'universo (2022) - Dipartimento di Matematica ``G.Peano'', Universit\`a degli Studi di Torino; this  work is done within the activity of the Gruppo Nazionale per la Fisica Matematica (GNFM).

%%------------------------Section-------------------------

\bibliography{short_note_s_rev}

\begin{thebibliography}{45}%
\makeatletter
\providecommand \@ifxundefined [1]{%
 \@ifx{#1\undefined}
}%
\providecommand \@ifnum [1]{%
 \ifnum #1\expandafter \@firstoftwo
 \else \expandafter \@secondoftwo
 \fi
}%
\providecommand \@ifx [1]{%
 \ifx #1\expandafter \@firstoftwo
 \else \expandafter \@secondoftwo
 \fi
}%
\providecommand \natexlab [1]{#1}%
\providecommand \enquote  [1]{``#1''}%
\providecommand \bibnamefont  [1]{#1}%
\providecommand \bibfnamefont [1]{#1}%
\providecommand \citenamefont [1]{#1}%
\providecommand \href@noop [0]{\@secondoftwo}%
\providecommand \href [0]{\begingroup \@sanitize@url \@href}%
\providecommand \@href[1]{\@@startlink{#1}\@@href}%
\providecommand \@@href[1]{\endgroup#1\@@endlink}%
\providecommand \@sanitize@url [0]{\catcode `\\12\catcode `\$12\catcode
  `\&12\catcode `\#12\catcode `\^12\catcode `\_12\catcode `\%12\relax}%
\providecommand \@@startlink[1]{}%
\providecommand \@@endlink[0]{}%
\providecommand \url  [0]{\begingroup\@sanitize@url \@url }%
\providecommand \@url [1]{\endgroup\@href {#1}{\urlprefix }}%
\providecommand \urlprefix  [0]{URL }%
\providecommand \Eprint [0]{\href }%
\providecommand \doibase [0]{http://dx.doi.org/}%
\providecommand \selectlanguage [0]{\@gobble}%
\providecommand \bibinfo  [0]{\@secondoftwo}%
\providecommand \bibfield  [0]{\@secondoftwo}%
\providecommand \translation [1]{[#1]}%
\providecommand \BibitemOpen [0]{}%
\providecommand \bibitemStop [0]{}%
\providecommand \bibitemNoStop [0]{.\EOS\space}%
\providecommand \EOS [0]{\spacefactor3000\relax}%
\providecommand \BibitemShut  [1]{\csname bibitem#1\endcsname}%
\let\auto@bib@innerbib\@empty
%</preamble>
\bibitem [{\citenamefont {Will}(2014)}]{Will:2014kxa}%
  \BibitemOpen
  \bibfield  {author} {\bibinfo {author} {\bibfnamefont {C.~M.}\ \bibnamefont
  {Will}},\ }\href {\doibase 10.12942/lrr-2014-4} {\bibfield  {journal}
  {\bibinfo  {journal} {Living Rev. Rel.}\ }\textbf {\bibinfo {volume} {17}},\
  \bibinfo {pages} {4} (\bibinfo {year} {2014})},\ \Eprint
  {http://arxiv.org/abs/1403.7377} {arXiv:1403.7377 [gr-qc]} \BibitemShut
  {NoStop}%
\bibitem [{\citenamefont {Ruggiero}\ and\ \citenamefont
  {Astesiano}(2023)}]{Ruggiero:2023ker}%
  \BibitemOpen
  \bibfield  {author} {\bibinfo {author} {\bibfnamefont {M.~L.}\ \bibnamefont
  {Ruggiero}}\ and\ \bibinfo {author} {\bibfnamefont {D.}~\bibnamefont
  {Astesiano}},\ }\href {\doibase 10.1088/2399-6528/ad08cf} {\bibfield
  {journal} {\bibinfo  {journal} {J. Phys. Comm.}\ }\textbf {\bibinfo {volume}
  {7}},\ \bibinfo {pages} {112001} (\bibinfo {year} {2023})},\ \Eprint
  {http://arxiv.org/abs/2304.02167} {arXiv:2304.02167 [gr-qc]} \BibitemShut
  {NoStop}%
\bibitem [{\citenamefont {Rindler}(2006)}]{rindler2006relativity}%
  \BibitemOpen
  \bibfield  {author} {\bibinfo {author} {\bibfnamefont {W.}~\bibnamefont
  {Rindler}},\ }\href@noop {} {\emph {\bibinfo {title} {Relativity: Special,
  General, and Cosmological}}}\ (\bibinfo  {publisher} {OUP Oxford},\ \bibinfo
  {year} {2006})\BibitemShut {NoStop}%
\bibitem [{\citenamefont {Geroch}(1971)}]{Geroch:1970nt}%
  \BibitemOpen
  \bibfield  {author} {\bibinfo {author} {\bibfnamefont {R.~P.}\ \bibnamefont
  {Geroch}},\ }\href {\doibase 10.1063/1.1665681} {\bibfield  {journal}
  {\bibinfo  {journal} {J. Math. Phys.}\ }\textbf {\bibinfo {volume} {12}},\
  \bibinfo {pages} {918} (\bibinfo {year} {1971})}\BibitemShut {NoStop}%
\bibitem [{\citenamefont {Geroch}(1972)}]{Geroch:1972yt}%
  \BibitemOpen
  \bibfield  {author} {\bibinfo {author} {\bibfnamefont {R.~P.}\ \bibnamefont
  {Geroch}},\ }\href {\doibase 10.1063/1.1665990} {\bibfield  {journal}
  {\bibinfo  {journal} {J. Math. Phys.}\ }\textbf {\bibinfo {volume} {13}},\
  \bibinfo {pages} {394} (\bibinfo {year} {1972})}\BibitemShut {NoStop}%
\bibitem [{\citenamefont {{Hansen}}\ and\ \citenamefont
  {{Winicour}}(1975)}]{HansenWinicour1}%
  \BibitemOpen
  \bibfield  {author} {\bibinfo {author} {\bibfnamefont {R.~O.}\ \bibnamefont
  {{Hansen}}}\ and\ \bibinfo {author} {\bibfnamefont {J.}~\bibnamefont
  {{Winicour}}},\ }\href {\doibase 10.1063/1.522608} {\bibfield  {journal}
  {\bibinfo  {journal} {Journal of Mathematical Physics}\ }\textbf {\bibinfo
  {volume} {16}},\ \bibinfo {pages} {804} (\bibinfo {year} {1975})}\BibitemShut
  {NoStop}%
\bibitem [{\citenamefont {Hansen}\ and\ \citenamefont
  {Winicour}(1977)}]{HansenWinicour2}%
  \BibitemOpen
  \bibfield  {author} {\bibinfo {author} {\bibfnamefont {R.~O.}\ \bibnamefont
  {Hansen}}\ and\ \bibinfo {author} {\bibfnamefont {J.}~\bibnamefont
  {Winicour}},\ }\href {\doibase 10.1063/1.523391} {\bibfield  {journal}
  {\bibinfo  {journal} {Journal of Mathematical Physics}\ }\textbf {\bibinfo
  {volume} {18}},\ \bibinfo {pages} {1206} (\bibinfo {year} {1977})},\ \Eprint
  {http://arxiv.org/abs/https://doi.org/10.1063/1.523391}
  {https://doi.org/10.1063/1.523391} \BibitemShut {NoStop}%
\bibitem [{\citenamefont {Winicour}(1975)}]{Winicour}%
  \BibitemOpen
  \bibfield  {author} {\bibinfo {author} {\bibfnamefont {J.}~\bibnamefont
  {Winicour}},\ }\href {\doibase 10.1063/1.522754} {\bibfield  {journal}
  {\bibinfo  {journal} {Journal of Mathematical Physics}\ }\textbf {\bibinfo
  {volume} {16}},\ \bibinfo {pages} {1806} (\bibinfo {year}
  {1975})}\BibitemShut {NoStop}%
\bibitem [{\citenamefont {Stephani}\ \emph {et~al.}(2003)\citenamefont
  {Stephani}, \citenamefont {Kramer}, \citenamefont {MacCallum}, \citenamefont
  {Hoenselaers},\ and\ \citenamefont {Herlt}}]{kramerstephani}%
  \BibitemOpen
  \bibfield  {author} {\bibinfo {author} {\bibfnamefont {H.}~\bibnamefont
  {Stephani}}, \bibinfo {author} {\bibfnamefont {D.}~\bibnamefont {Kramer}},
  \bibinfo {author} {\bibfnamefont {M.}~\bibnamefont {MacCallum}}, \bibinfo
  {author} {\bibfnamefont {C.}~\bibnamefont {Hoenselaers}}, \ and\ \bibinfo
  {author} {\bibfnamefont {E.}~\bibnamefont {Herlt}},\ }\href {\doibase
  10.1017/CBO9780511535185} {\emph {\bibinfo {title} {Exact Solutions of
  Einstein's Field Equations}}},\ \bibinfo {edition} {2nd}\ ed.,\ Cambridge
  Monographs on Mathematical Physics\ (\bibinfo  {publisher} {Cambridge
  University Press},\ \bibinfo {year} {2003})\BibitemShut {NoStop}%
\bibitem [{\citenamefont {{Bardeen}}\ \emph {et~al.}(1972)\citenamefont
  {{Bardeen}}, \citenamefont {{Press}},\ and\ \citenamefont
  {{Teukolsky}}}]{zamo}%
  \BibitemOpen
  \bibfield  {author} {\bibinfo {author} {\bibfnamefont {J.~M.}\ \bibnamefont
  {{Bardeen}}}, \bibinfo {author} {\bibfnamefont {W.~H.}\ \bibnamefont
  {{Press}}}, \ and\ \bibinfo {author} {\bibfnamefont {S.~A.}\ \bibnamefont
  {{Teukolsky}}},\ }\href {\doibase 10.1086/151796} {\bibfield  {journal}
  {\bibinfo  {journal} {The Astrophysical Journal}\ }\textbf {\bibinfo {volume}
  {178}},\ \bibinfo {pages} {347} (\bibinfo {year} {1972})}\BibitemShut
  {NoStop}%
\bibitem [{\citenamefont {Astesiano}\ and\ \citenamefont
  {Ruggiero}(2022{\natexlab{a}})}]{astesiano}%
  \BibitemOpen
  \bibfield  {author} {\bibinfo {author} {\bibfnamefont {D.}~\bibnamefont
  {Astesiano}}\ and\ \bibinfo {author} {\bibfnamefont {M.~L.}\ \bibnamefont
  {Ruggiero}},\ }\href {\doibase 10.1103/PhysRevD.106.044061} {\bibfield
  {journal} {\bibinfo  {journal} {Phys. Rev. D}\ }\textbf {\bibinfo {volume}
  {106}},\ \bibinfo {pages} {044061} (\bibinfo {year}
  {2022}{\natexlab{a}})}\BibitemShut {NoStop}%
\bibitem [{\citenamefont {Ruggiero}\ and\ \citenamefont
  {Tartaglia}(2002)}]{Ruggiero:2002hz}%
  \BibitemOpen
  \bibfield  {author} {\bibinfo {author} {\bibfnamefont {M.~L.}\ \bibnamefont
  {Ruggiero}}\ and\ \bibinfo {author} {\bibfnamefont {A.}~\bibnamefont
  {Tartaglia}},\ }\href@noop {} {\bibfield  {journal} {\bibinfo  {journal}
  {Nuovo Cim.}\ }\textbf {\bibinfo {volume} {B117}},\ \bibinfo {pages} {743}
  (\bibinfo {year} {2002})},\ \Eprint {http://arxiv.org/abs/gr-qc/0207065}
  {arXiv:gr-qc/0207065 [gr-qc]} \BibitemShut {NoStop}%
%%CITATION = GR-QC/0207065;%%
\bibitem [{\citenamefont {{Mashhoon}}(2007)}]{mashhoon03}%
  \BibitemOpen
  \bibfield  {author} {\bibinfo {author} {\bibfnamefont {B.}~\bibnamefont
  {{Mashhoon}}},\ }in\ \href@noop {} {\emph {\bibinfo {booktitle} {The
  measurement of gravitomagnetism: a challenging enterprise}}},\ \bibinfo
  {editor} {edited by\ \bibinfo {editor} {\bibfnamefont {L.}~\bibnamefont
  {Iorio}}}\ (\bibinfo  {publisher} {NOVA publishers},\ \bibinfo {year}
  {2007})\ \Eprint {http://arxiv.org/abs/gr-qc/0311030} {arXiv:gr-qc/0311030}
  \BibitemShut {NoStop}%
\bibitem [{\citenamefont {Bonnor}(1977)}]{bonnor1977rotating}%
  \BibitemOpen
  \bibfield  {author} {\bibinfo {author} {\bibfnamefont {W.}~\bibnamefont
  {Bonnor}},\ }\href {\doibase 10.1088/0305-4470/10/10/004} {\bibfield
  {journal} {\bibinfo  {journal} {Journal of Physics A: Mathematical and
  General}\ }\textbf {\bibinfo {volume} {10}},\ \bibinfo {pages} {1673}
  (\bibinfo {year} {1977})}\BibitemShut {NoStop}%
\bibitem [{\citenamefont {Grad}\ and\ \citenamefont
  {Rubin}(1958)}]{grad1958hydromagnetic}%
  \BibitemOpen
  \bibfield  {author} {\bibinfo {author} {\bibfnamefont {H.}~\bibnamefont
  {Grad}}\ and\ \bibinfo {author} {\bibfnamefont {H.}~\bibnamefont {Rubin}},\
  }\href@noop {} {\bibfield  {journal} {\bibinfo  {journal} {Journal of Nuclear
  Energy (1954)}\ }\textbf {\bibinfo {volume} {7}},\ \bibinfo {pages} {284}
  (\bibinfo {year} {1958})}\BibitemShut {NoStop}%
\bibitem [{\citenamefont
  {Shafranov}(1958)}]{shafranov1958magnetohydrodynamical}%
  \BibitemOpen
  \bibfield  {author} {\bibinfo {author} {\bibfnamefont {V.}~\bibnamefont
  {Shafranov}},\ }\href@noop {} {\bibfield  {journal} {\bibinfo  {journal}
  {Soviet Physics JETP}\ }\textbf {\bibinfo {volume} {6}},\ \bibinfo {pages}
  {1013} (\bibinfo {year} {1958})}\BibitemShut {NoStop}%
\bibitem [{\citenamefont {Astesiano}\ and\ \citenamefont
  {Ruggiero}(2022{\natexlab{b}})}]{Astesiano:2022ghr}%
  \BibitemOpen
  \bibfield  {author} {\bibinfo {author} {\bibfnamefont {D.}~\bibnamefont
  {Astesiano}}\ and\ \bibinfo {author} {\bibfnamefont {M.~L.}\ \bibnamefont
  {Ruggiero}},\ }\href {\doibase 10.1103/PhysRevD.106.L121501} {\bibfield
  {journal} {\bibinfo  {journal} {Phys. Rev. D}\ }\textbf {\bibinfo {volume}
  {106}},\ \bibinfo {pages} {L121501} (\bibinfo {year} {2022}{\natexlab{b}})},\
  \Eprint {http://arxiv.org/abs/2211.11815} {arXiv:2211.11815 [gr-qc]}
  \BibitemShut {NoStop}%
\bibitem [{\citenamefont {Astesiano}\ \emph {et~al.}(2022)\citenamefont
  {Astesiano}, \citenamefont {Cacciatori}, \citenamefont {Gorini},\ and\
  \citenamefont {Re}}]{Astesiano:2021ren}%
  \BibitemOpen
  \bibfield  {author} {\bibinfo {author} {\bibfnamefont {D.}~\bibnamefont
  {Astesiano}}, \bibinfo {author} {\bibfnamefont {S.~L.}\ \bibnamefont
  {Cacciatori}}, \bibinfo {author} {\bibfnamefont {V.}~\bibnamefont {Gorini}},
  \ and\ \bibinfo {author} {\bibfnamefont {F.}~\bibnamefont {Re}},\ }\href
  {\doibase 10.1140/epjc/s10052-022-10506-7} {\bibfield  {journal} {\bibinfo
  {journal} {Eur. Phys. J. C}\ }\textbf {\bibinfo {volume} {82}},\ \bibinfo
  {pages} {554} (\bibinfo {year} {2022})},\ \Eprint
  {http://arxiv.org/abs/2106.12818} {arXiv:2106.12818 [gr-qc]} \BibitemShut
  {NoStop}%
\bibitem [{\citenamefont {{Crisanti}}(2019)}]{crisanti}%
  \BibitemOpen
  \bibfield  {author} {\bibinfo {author} {\bibfnamefont {F.}~\bibnamefont
  {{Crisanti}}},\ }\href {\doibase 10.1017/S0022377819000175} {\bibfield
  {journal} {\bibinfo  {journal} {Journal of Plasma Physics}\ }\textbf
  {\bibinfo {volume} {85}},\ \bibinfo {eid} {905850210} (\bibinfo {year}
  {2019})}\BibitemShut {NoStop}%
\bibitem [{\citenamefont {Lupica}\ \emph {et~al.}(2021)\citenamefont {Lupica},
  \citenamefont {Cesarano}, \citenamefont {Crisanti},\ and\ \citenamefont
  {Ishkhanyan}}]{cesarano}%
  \BibitemOpen
  \bibfield  {author} {\bibinfo {author} {\bibfnamefont {A.}~\bibnamefont
  {Lupica}}, \bibinfo {author} {\bibfnamefont {C.}~\bibnamefont {Cesarano}},
  \bibinfo {author} {\bibfnamefont {F.}~\bibnamefont {Crisanti}}, \ and\
  \bibinfo {author} {\bibfnamefont {A.}~\bibnamefont {Ishkhanyan}},\ }\href
  {\doibase 10.3390/math9243316} {\bibfield  {journal} {\bibinfo  {journal}
  {Mathematics}\ }\textbf {\bibinfo {volume} {9}} (\bibinfo {year} {2021}),\
  10.3390/math9243316}\BibitemShut {NoStop}%
\bibitem [{\citenamefont {{Sob{\v{e}}hart}}(1990)}]{multipole1}%
  \BibitemOpen
  \bibfield  {author} {\bibinfo {author} {\bibfnamefont {J.~R.}\ \bibnamefont
  {{Sob{\v{e}}hart}}},\ }\href {\doibase 10.1063/1.859532} {\bibfield
  {journal} {\bibinfo  {journal} {Physics of Fluids B}\ }\textbf {\bibinfo
  {volume} {2}},\ \bibinfo {pages} {222} (\bibinfo {year} {1990})}\BibitemShut
  {NoStop}%
\bibitem [{\citenamefont {Reusch}\ and\ \citenamefont
  {Neilson}(1984)}]{multipole2}%
  \BibitemOpen
  \bibfield  {author} {\bibinfo {author} {\bibfnamefont {M.~F.}\ \bibnamefont
  {Reusch}}\ and\ \bibinfo {author} {\bibfnamefont {G.~H.}\ \bibnamefont
  {Neilson}},\ }\href@noop {} {\emph {\bibinfo {title} {Finite order polynomial
  moment solutions of the homogeneous Grad-Shafranov equation}}},\ \bibinfo
  {type} {Tech. Rep.}\ (\bibinfo  {institution} {Princeton Univ., NJ (USA).
  Plasma Physics Lab.; Oak Ridge National Lab., TN~{\ldots}},\ \bibinfo {year}
  {1984})\BibitemShut {NoStop}%
\bibitem [{\citenamefont {Balasin}\ and\ \citenamefont
  {Grumiller}(2008)}]{Balasin:2006cg}%
  \BibitemOpen
  \bibfield  {author} {\bibinfo {author} {\bibfnamefont {H.}~\bibnamefont
  {Balasin}}\ and\ \bibinfo {author} {\bibfnamefont {D.}~\bibnamefont
  {Grumiller}},\ }\href {\doibase 10.1142/S0218271808012140} {\bibfield
  {journal} {\bibinfo  {journal} {Int. J. Mod. Phys. D}\ }\textbf {\bibinfo
  {volume} {17}},\ \bibinfo {pages} {475} (\bibinfo {year} {2008})},\ \Eprint
  {http://arxiv.org/abs/astro-ph/0602519} {arXiv:astro-ph/0602519} \BibitemShut
  {NoStop}%
\bibitem [{\citenamefont {Astesiano}(2022)}]{Astesiano:2022gph}%
  \BibitemOpen
  \bibfield  {author} {\bibinfo {author} {\bibfnamefont {D.}~\bibnamefont
  {Astesiano}},\ }\href {\doibase 10.1007/s10714-022-02947-y} {\bibfield
  {journal} {\bibinfo  {journal} {Gen. Rel. Grav.}\ }\textbf {\bibinfo {volume}
  {54}},\ \bibinfo {pages} {63} (\bibinfo {year} {2022})},\ \Eprint
  {http://arxiv.org/abs/2201.03959} {arXiv:2201.03959 [gr-qc]} \BibitemShut
  {NoStop}%
\bibitem [{\citenamefont {Costa}\ \emph {et~al.}(2023)\citenamefont {Costa},
  \citenamefont {Nat\'ario}, \citenamefont {Frutos-Alfaro},\ and\ \citenamefont
  {Soffel}}]{costaBG}%
  \BibitemOpen
  \bibfield  {author} {\bibinfo {author} {\bibfnamefont {L.~F.~O.}\
  \bibnamefont {Costa}}, \bibinfo {author} {\bibfnamefont {J.}~\bibnamefont
  {Nat\'ario}}, \bibinfo {author} {\bibfnamefont {F.}~\bibnamefont
  {Frutos-Alfaro}}, \ and\ \bibinfo {author} {\bibfnamefont {M.}~\bibnamefont
  {Soffel}},\ }\href@noop {} {\  (\bibinfo {year} {2023})},\ \Eprint
  {http://arxiv.org/abs/2303.17516} {arXiv:2303.17516 [gr-qc]} \BibitemShut
  {NoStop}%
\bibitem [{\citenamefont {{Jefremov}}\ and\ \citenamefont
  {{Perlick}}(2016)}]{perlick}%
  \BibitemOpen
  \bibfield  {author} {\bibinfo {author} {\bibfnamefont {P.~I.}\ \bibnamefont
  {{Jefremov}}}\ and\ \bibinfo {author} {\bibfnamefont {V.}~\bibnamefont
  {{Perlick}}},\ }\href {\doibase 10.1088/0264-9381/33/24/245014} {\bibfield
  {journal} {\bibinfo  {journal} {Classical and Quantum Gravity}\ }\textbf
  {\bibinfo {volume} {33}},\ \bibinfo {eid} {245014} (\bibinfo {year}
  {2016})},\ \Eprint {http://arxiv.org/abs/1608.06218} {arXiv:1608.06218
  [gr-qc]} \BibitemShut {NoStop}%
\bibitem [{\citenamefont {{Reina}}\ and\ \citenamefont
  {{Treves}}(1976)}]{treves}%
  \BibitemOpen
  \bibfield  {author} {\bibinfo {author} {\bibfnamefont {C.}~\bibnamefont
  {{Reina}}}\ and\ \bibinfo {author} {\bibfnamefont {A.}~\bibnamefont
  {{Treves}}},\ }\href {\doibase 10.1007/BF00778761} {\bibfield  {journal}
  {\bibinfo  {journal} {General Relativity and Gravitation}\ }\textbf {\bibinfo
  {volume} {7}},\ \bibinfo {pages} {817} (\bibinfo {year} {1976})}\BibitemShut
  {NoStop}%
\bibitem [{\citenamefont {{Bonnor}}(1992)}]{bonnor_exact}%
  \BibitemOpen
  \bibfield  {author} {\bibinfo {author} {\bibfnamefont {W.~B.}\ \bibnamefont
  {{Bonnor}}},\ }\href {\doibase 10.1007/BF00760137} {\bibfield  {journal}
  {\bibinfo  {journal} {General Relativity and Gravitation}\ }\textbf {\bibinfo
  {volume} {24}},\ \bibinfo {pages} {551} (\bibinfo {year} {1992})}\BibitemShut
  {NoStop}%
\bibitem [{\citenamefont {{Neugebauer}}\ and\ \citenamefont
  {{Meinel}}(1993)}]{Neugebauer93}%
  \BibitemOpen
  \bibfield  {author} {\bibinfo {author} {\bibfnamefont {G.}~\bibnamefont
  {{Neugebauer}}}\ and\ \bibinfo {author} {\bibfnamefont {R.}~\bibnamefont
  {{Meinel}}},\ }\href {\doibase 10.1086/187005} {\bibfield  {journal}
  {\bibinfo  {journal} {\apjl}\ }\textbf {\bibinfo {volume} {414}},\ \bibinfo
  {pages} {L97} (\bibinfo {year} {1993})}\BibitemShut {NoStop}%
\bibitem [{\citenamefont {Neugebauer}\ \emph {et~al.}(1996)\citenamefont
  {Neugebauer}, \citenamefont {Kleinwachter},\ and\ \citenamefont
  {Meinel}}]{Neugebauer:1996sz}%
  \BibitemOpen
  \bibfield  {author} {\bibinfo {author} {\bibfnamefont {G.}~\bibnamefont
  {Neugebauer}}, \bibinfo {author} {\bibfnamefont {A.}~\bibnamefont
  {Kleinwachter}}, \ and\ \bibinfo {author} {\bibfnamefont {R.}~\bibnamefont
  {Meinel}},\ }\href@noop {} {\bibfield  {journal} {\bibinfo  {journal} {Helv.
  Phys. Acta}\ }\textbf {\bibinfo {volume} {69}},\ \bibinfo {pages} {472}
  (\bibinfo {year} {1996})},\ \Eprint {http://arxiv.org/abs/gr-qc/0301107}
  {arXiv:gr-qc/0301107} \BibitemShut {NoStop}%
\bibitem [{\citenamefont {Ilyas}\ \emph {et~al.}(2017)\citenamefont {Ilyas},
  \citenamefont {Yang}, \citenamefont {Malafarina},\ and\ \citenamefont
  {Bambi}}]{bambi}%
  \BibitemOpen
  \bibfield  {author} {\bibinfo {author} {\bibfnamefont {B.}~\bibnamefont
  {Ilyas}}, \bibinfo {author} {\bibfnamefont {J.}~\bibnamefont {Yang}},
  \bibinfo {author} {\bibfnamefont {D.}~\bibnamefont {Malafarina}}, \ and\
  \bibinfo {author} {\bibfnamefont {C.}~\bibnamefont {Bambi}},\ }\href
  {\doibase 10.1140/epjc/s10052-017-5014-3} {\bibfield  {journal} {\bibinfo
  {journal} {Eur. Phys. J. C}\ }\textbf {\bibinfo {volume} {77}},\ \bibinfo
  {pages} {461} (\bibinfo {year} {2017})},\ \Eprint
  {http://arxiv.org/abs/1611.03972} {arXiv:1611.03972 [gr-qc]} \BibitemShut
  {NoStop}%
\bibitem [{\citenamefont {Astesiano}\ \emph {et~al.}(2023)\citenamefont
  {Astesiano}, \citenamefont {Bini}, \citenamefont {Geralico},\ and\
  \citenamefont {Ruggiero}}]{Astesiano:2023emg}%
  \BibitemOpen
  \bibfield  {author} {\bibinfo {author} {\bibfnamefont {D.}~\bibnamefont
  {Astesiano}}, \bibinfo {author} {\bibfnamefont {D.}~\bibnamefont {Bini}},
  \bibinfo {author} {\bibfnamefont {A.}~\bibnamefont {Geralico}}, \ and\
  \bibinfo {author} {\bibfnamefont {M.~L.}\ \bibnamefont {Ruggiero}},\
  }\href@noop {} {\  (\bibinfo {year} {2023})},\ \Eprint
  {http://arxiv.org/abs/2310.04157} {arXiv:2310.04157 [gr-qc]} \BibitemShut
  {NoStop}%
\bibitem [{\citenamefont {{Binney}}\ and\ \citenamefont
  {{Tremaine}}(1988)}]{binneytremaine}%
  \BibitemOpen
  \bibfield  {author} {\bibinfo {author} {\bibfnamefont {J.}~\bibnamefont
  {{Binney}}}\ and\ \bibinfo {author} {\bibfnamefont {S.}~\bibnamefont
  {{Tremaine}}},\ }\href@noop {} {\emph {\bibinfo {title} {{Galactic
  Dynamics}}}}\ (\bibinfo {year} {1988})\BibitemShut {NoStop}%
\bibitem [{\citenamefont {{Rendall}}(1992)}]{rendall}%
  \BibitemOpen
  \bibfield  {author} {\bibinfo {author} {\bibfnamefont {A.~D.}\ \bibnamefont
  {{Rendall}}},\ }\href {\doibase 10.1088/0264-9381/9/8/005} {\bibfield
  {journal} {\bibinfo  {journal} {Classical and Quantum Gravity}\ }\textbf
  {\bibinfo {volume} {9}},\ \bibinfo {pages} {L99} (\bibinfo {year}
  {1992})}\BibitemShut {NoStop}%
\bibitem [{\citenamefont {{Ludwig}}(2021)}]{Ludwig}%
  \BibitemOpen
  \bibfield  {author} {\bibinfo {author} {\bibfnamefont {G.~O.}\ \bibnamefont
  {{Ludwig}}},\ }\href {\doibase 10.1140/epjc/s10052-021-08967-3} {\bibfield
  {journal} {\bibinfo  {journal} {European Physical Journal C}\ }\textbf
  {\bibinfo {volume} {81}},\ \bibinfo {eid} {186} (\bibinfo {year}
  {2021})}\BibitemShut {NoStop}%
\bibitem [{\citenamefont {Ruggiero}\ \emph {et~al.}(2022)\citenamefont
  {Ruggiero}, \citenamefont {Ortolan},\ and\ \citenamefont
  {Speake}}]{Ruggiero:2021lpf}%
  \BibitemOpen
  \bibfield  {author} {\bibinfo {author} {\bibfnamefont {M.~L.}\ \bibnamefont
  {Ruggiero}}, \bibinfo {author} {\bibfnamefont {A.}~\bibnamefont {Ortolan}}, \
  and\ \bibinfo {author} {\bibfnamefont {C.~C.}\ \bibnamefont {Speake}},\
  }\href {\doibase 10.1088/1361-6382/ac9949} {\bibfield  {journal} {\bibinfo
  {journal} {Class. Quant. Grav.}\ }\textbf {\bibinfo {volume} {39}},\ \bibinfo
  {pages} {225015} (\bibinfo {year} {2022})},\ \Eprint
  {http://arxiv.org/abs/2112.08290} {arXiv:2112.08290 [gr-qc]} \BibitemShut
  {NoStop}%
\bibitem [{\citenamefont {Ciotti}(2022)}]{Ciotti_2022}%
  \BibitemOpen
  \bibfield  {author} {\bibinfo {author} {\bibfnamefont {L.}~\bibnamefont
  {Ciotti}},\ }\href {\doibase 10.3847/1538-4357/ac82b3} {\bibfield  {journal}
  {\bibinfo  {journal} {The Astrophysical Journal}\ }\textbf {\bibinfo {volume}
  {936}},\ \bibinfo {pages} {180} (\bibinfo {year} {2022})}\BibitemShut
  {NoStop}%
\bibitem [{\citenamefont {Govaerts}(2023)}]{jan}%
  \BibitemOpen
  \bibfield  {author} {\bibinfo {author} {\bibfnamefont {J.}~\bibnamefont
  {Govaerts}},\ }\href {\doibase 10.1088/1361-6382/acc22d} {\  (\bibinfo {year}
  {2023}),\ 10.1088/1361-6382/acc22d},\ \Eprint
  {http://arxiv.org/abs/2303.01386} {arXiv:2303.01386 [gr-qc]} \BibitemShut
  {NoStop}%
\bibitem [{\citenamefont {Rubin}\ \emph {et~al.}(1978)\citenamefont {Rubin},
  \citenamefont {Ford~Jr},\ and\ \citenamefont {Thonnard}}]{rubin1978extended}%
  \BibitemOpen
  \bibfield  {author} {\bibinfo {author} {\bibfnamefont {V.~C.}\ \bibnamefont
  {Rubin}}, \bibinfo {author} {\bibfnamefont {W.~K.}\ \bibnamefont {Ford~Jr}},
  \ and\ \bibinfo {author} {\bibfnamefont {N.}~\bibnamefont {Thonnard}},\
  }\href@noop {} {\bibfield  {journal} {\bibinfo  {journal} {The Astrophysical
  Journal}\ }\textbf {\bibinfo {volume} {225}},\ \bibinfo {pages} {L107}
  (\bibinfo {year} {1978})}\BibitemShut {NoStop}%
\bibitem [{\citenamefont {Sofue}\ and\ \citenamefont
  {Rubin}(2001)}]{sofue2001rotation}%
  \BibitemOpen
  \bibfield  {author} {\bibinfo {author} {\bibfnamefont {Y.}~\bibnamefont
  {Sofue}}\ and\ \bibinfo {author} {\bibfnamefont {V.}~\bibnamefont {Rubin}},\
  }\href@noop {} {\bibfield  {journal} {\bibinfo  {journal} {Annual Review of
  Astronomy and Astrophysics}\ }\textbf {\bibinfo {volume} {39}},\ \bibinfo
  {pages} {137} (\bibinfo {year} {2001})}\BibitemShut {NoStop}%
\bibitem [{\citenamefont {{Sofue}}(2020)}]{sofuegalaxy}%
  \BibitemOpen
  \bibfield  {author} {\bibinfo {author} {\bibfnamefont {Y.}~\bibnamefont
  {{Sofue}}},\ }\href {\doibase 10.3390/galaxies8020037} {\bibfield  {journal}
  {\bibinfo  {journal} {Galaxies}\ }\textbf {\bibinfo {volume} {8}},\ \bibinfo
  {pages} {37} (\bibinfo {year} {2020})},\ \Eprint
  {http://arxiv.org/abs/2004.11688} {arXiv:2004.11688 [astro-ph.GA]}
  \BibitemShut {NoStop}%
\bibitem [{\citenamefont {{Miller}}(1973)}]{miller}%
  \BibitemOpen
  \bibfield  {author} {\bibinfo {author} {\bibfnamefont {J.~G.}\ \bibnamefont
  {{Miller}}},\ }\href {\doibase 10.1063/1.1666343} {\bibfield  {journal}
  {\bibinfo  {journal} {Journal of Mathematical Physics}\ }\textbf {\bibinfo
  {volume} {14}},\ \bibinfo {pages} {486} (\bibinfo {year} {1973})}\BibitemShut
  {NoStop}%
\bibitem [{\citenamefont {{Chakraborty}}\ and\ \citenamefont
  {{Bhattacharyya}}(2019)}]{attuale1}%
  \BibitemOpen
  \bibfield  {author} {\bibinfo {author} {\bibfnamefont {C.}~\bibnamefont
  {{Chakraborty}}}\ and\ \bibinfo {author} {\bibfnamefont {S.}~\bibnamefont
  {{Bhattacharyya}}},\ }\href {\doibase 10.1088/1475-7516/2019/05/034}
  {\bibfield  {journal} {\bibinfo  {journal} {\jcap}\ }\textbf {\bibinfo
  {volume} {2019}},\ \bibinfo {eid} {034} (\bibinfo {year} {2019})},\ \Eprint
  {http://arxiv.org/abs/1901.04233} {arXiv:1901.04233 [astro-ph.HE]}
  \BibitemShut {NoStop}%
\bibitem [{\citenamefont {{Chakraborty}}\ and\ \citenamefont
  {{Bhattacharyya}}(2018)}]{attuale2}%
  \BibitemOpen
  \bibfield  {author} {\bibinfo {author} {\bibfnamefont {C.}~\bibnamefont
  {{Chakraborty}}}\ and\ \bibinfo {author} {\bibfnamefont {S.}~\bibnamefont
  {{Bhattacharyya}}},\ }\href {\doibase 10.1103/PhysRevD.98.043021} {\bibfield
  {journal} {\bibinfo  {journal} {\prd}\ }\textbf {\bibinfo {volume} {98}},\
  \bibinfo {eid} {043021} (\bibinfo {year} {2018})},\ \Eprint
  {http://arxiv.org/abs/1712.01156} {arXiv:1712.01156 [astro-ph.HE]}
  \BibitemShut {NoStop}%
\bibitem [{\citenamefont {Akiyama}\ \emph {et~al.}(2019)\citenamefont {Akiyama}
  \emph {et~al.}}]{ETBH}%
  \BibitemOpen
  \bibfield  {author} {\bibinfo {author} {\bibfnamefont {K.}~\bibnamefont
  {Akiyama}} \emph {et~al.} (\bibinfo {collaboration} {Event Horizon
  Telescope}),\ }\href {\doibase 10.3847/2041-8213/ab0ec7} {\bibfield
  {journal} {\bibinfo  {journal} {Astrophys. J. Lett.}\ }\textbf {\bibinfo
  {volume} {875}},\ \bibinfo {pages} {L1} (\bibinfo {year} {2019})},\ \Eprint
  {http://arxiv.org/abs/1906.11238} {arXiv:1906.11238 [astro-ph.GA]}
  \BibitemShut {NoStop}%
\end{thebibliography}%

\end{document}